# Title: Aerosol invigoration of atmospheric convection through increases in humidity

**Authors:** Tristan H. Abbott[1]*, Timothy W. Cronin[1]

**Affiliations:**

[1]Department of Earth, Atmospheric, and Planetary Sciences, Massachusetts Institute of Technology, 77 Massachusetts Ave., Cambridge 02139, USA.

*Correspondence to: thabbott@mit.edu

**One Sentence Summary:** Idealized simulations show that aerosols enhance tropical thunderstorm activity by increasing humidity outside clouds.

**Abstract:** Cloud-aerosol interactions remain a major obstacle to understanding climate and severe weather. Observations suggest that aerosols enhance tropical thunderstorm activity; past research, motivated by the importance of understanding aerosol impacts on clouds, has proposed several mechanisms that could explain that observed link. Here, we show that high-resolution atmospheric simulations can reproduce the observed link between aerosols and convection. However, we also show that previously proposed mechanisms are unable to explain the invigoration. Examining underlying processes reveals that, in our simulations, high aerosol concentrations increase environmental humidity by producing clouds that mix more condensed water into the surrounding air. In turn, higher humidity favors large-scale ascent and stronger convection. Our results provide a physical reason to expect invigorated thunderstorms in high-aerosol regions of the tropics.

**Main Text:** Observations suggest cloud-aerosol interactions play a significant role in setting the frequency and intensity of atmospheric deep convection. Many studies have found increases in cloud top height and cloud cover coincident with increases in aerosol loading *(1-5)*; additionally, lightning flash rates are consistently higher in high-aerosol regions of the tropics, including continents, islands, and ship tracks *(6-8)*. These observations indicate that high aerosol concentrations may trigger a chain of processes that ultimately increases the number or strength of convective updrafts—which we refer to throughout this paper as "microphysical invigoration". A deeper understanding of microphysical invigoration could enhance understanding of severe weather risks and of climatically-important cloud-aerosol interactions *(9-12)*, and is particularly relevant because human activity is a major aerosol source.

Past work has proposed two mechanisms by which aerosol concentrations could invigorate convection. The first mechanism relies on a "cold-phase" pathway: higher concentrations of cloud condensation nuclei (particles onto which liquid cloud droplets condense) suppress rain in shallow clouds, allowing clouds to loft more condensate through the freezing level, increasing the latent heat released when cloud water freezes, and enhancing buoyancy once enough condensate precipitates out *(13)*. The second mechanism relies on a "warm-phase" pathway: higher aerosol concentrations reduce supersaturation in liquid clouds, increasing latent heat release through additional condensation of water vapor *(14)*.

Here, we use idealized high-resolution simulations with the System for Atmospheric Modeling (SAM) *(15)*, scale analysis, and plume model calculations to describe a novel "humidity-entrainment" invigoration mechanism that is distinct from both the cold- and warm-phase mechanisms. Unlike the cold- and warm-phase mechanisms, which consider aerosol-induced changes in cloud processes independently from changes in the surrounding clear-air environment, the humidity-entrainment mechanism relies specifically on cloud-environment feedbacks. The two key ingredients are (1) an increase in environmental humidity in response to higher aerosol concentrations in clouds, and (2) an increase in large-scale ascent in response to increased environmental humidity.

We represent changes in aerosol abundance in our simulations by varying a prescribed liquid cloud droplet number concentration ($N_c$) in SAM's cloud microphysics scheme *(16)* from 50 cm$^{-3}$ (characteristic of pristine maritime environments) to 800 cm$^{-3}$ (characteristic of polluted continental environments) *(17)*. Because larger $N_c$ inhibits rain formation in liquid clouds *(18)*, the cold-phase invigoration mechanism could plausibly invigorate convection in simulations with higher $N_c$. However, we configure the microphysics scheme to allow no supersaturation in liquid clouds, precluding the operation of the warm-phase invigoration mechanism.

We first consider simulations that include a parameterization of large-scale dynamics based on the weak temperature gradient (WTG) approximation *(19, 20)*. The WTG parameterization diagnoses large-scale vertical motion that relaxes domain-average temperature profiles toward a reference profile and generates large-scale moisture convergence. We configure WTG-constrained simulations to mimic a localized aerosol anomaly embedded within a large-scale low-aerosol environment, but with otherwise identical boundary conditions, by taking the reference temperature profile from an $N_c$ = 50 cm$^{-3}$ simulation of radiative-convective equilibrium (RCE), where the environment evolves freely until reaching an equilibrium where heating by convection balances cooling by radiation.

Larger $N_c$ increases large-scale ascent by up to 4 cm s$^{-1}$ around 8 km (Fig. 1A). Because of constraints imposed by WTG dynamics, increasing $N_c$ leaves environmental temperature profiles nearly unaltered (Fig. 1B). Increases in large-scale ascent with $N_c$ are accompanied by increases in the upward mass flux in clouds (Fig. 1C), consistent with constraints imposed by steady-state energy and mass balances (Supplementary Text), as well as by increases in humidity and precipitation (Fig. 1D, S1). High-percentile vertical velocities also increase with $N_c$, by up to 100% in the upper troposphere (Fig. 1E). The increase in high-percentile vertical velocity is accompanied by greater areal coverage of strong convective cores, but not by greater average updraft speeds within strong convective cores (Fig. S2). This suggests that increased high-percentile vertical velocities are linked to increased frequency rather than increased speed of strong updrafts. Larger cloud mass fluxes and high-percentile vertical velocities are both signatures of convective invigoration, and these results—invigoration in simulations with a localized aerosol anomaly—are consistent with results from other modeling studies that have represented aerosol anomalies using spatially inhomogeneous RCE simulations *(21, 22)* rather than by coupling domain-wide aerosol changes to parameterized large-scale dynamics.

If we remove the WTG parameterization and instead run RCE simulations with varied $N_c$ *(20)*, the large-scale vertical velocity is constrained to vanish, and increasing $N_c$ no longer leads to substantial changes in cloud mass fluxes or high-percentile vertical velocities (Fig. 2C,D). Instead, heating that would trigger large-scale ascent under WTG instead leads to increases in atmospheric temperatures (Fig. 2A,B). Although RCE simulations lack invigoration, examining the processes that warm the atmosphere in RCE simulations provides insight into invigoration in WTG simulations. Additionally, focusing on understanding those processes in RCE simulations allows us to link changes in microphysics and atmospheric heating without the complexity introduced by parameterized large-scale circulations.

Because the cold-phase invigoration mechanism relies on the latent heat of fusion, we test whether it is responsible for the warmer troposphere in high-$N_c$ RCE simulations by conducting mechanism-denial experiments with the latent heat of fusion $L_f$ set to 0. While a slightly deeper layer of the troposphere is warmed when the latent heat of fusion is non-zero, we find that differences between low- and high-$N_c$ temperature profiles with $L_f = 0$ are almost identical to the default simulations (Fig. 3A). The persistence of the temperature differences provides strong evidence that the cold-phase mechanism is not responsible for warming the upper troposphere in high-$N_c$ simulations of RCE.

If not the latent heat of fusion, then what produces a warmer middle and upper troposphere in simulations with higher $N_c$? Humidity changes are one possibility. Previous work has shown that updraft temperatures are closely linked, through entrainment, to environmental relative humidity *(24)*. Moreover, scale analysis (Supplementary Text) suggests that parcel temperatures might increase by as much as 1 K per 1% change in relative humidity, whereas changes in the amount of frozen condensate are unlikely to change parcel temperatures by more than 1 K total.

A simple plume model, which links tropospheric warming to increased humidity in RCE simulations, provides evidence that changes in humidity are responsible for warming the troposphere at high $N_c$ *(20)*. Mean specific humidity profiles in RCE simulations increase with $N_c$ between 2.5 and 7.5 km (Fig. 3B). Because updrafts are cooled less by entrainment in a moister environment, the plume model predicts warmer temperature profiles in high-$N_c$ simulations above about 3 to 4 km, approximately coincident with the level where simulated temperature differences first become large (Fig. 3C). With an entrainment rate of 1 km$^{-1}$, the plume model also reproduces the magnitude of temperature differences in the upper troposphere, although this is sensitive to the choice of entrainment rate. Overall, our mechanism-denial experiments, scale analysis, and plume calculations all point to changes in atmospheric humidity as the driver of tropospheric warming at high $N_c$.

But why does humidity itself change in response to aerosol increases? In RCE, environmental humidity is set by a balance between moistening by air detrained from clouds and drying by clear-air subsidence *(25,26)*. By inhibiting the formation of rain in liquid clouds *(18)*, increases in aerosol concentrations may increase the mass of condensate in air detrained below the freezing level, which in turn could act to increase tropospheric relative humidity by increasing detrainment moistening *(26)*. High-$N_c$ RCE simulations show larger condensate concentrations and higher cloud evaporation rates near levels where humidity increases (Fig. 4); both features are consistent with higher humidity driven by aerosol-induced changes in detrainment moistening.

To summarize: our analysis of RCE simulations points to a causal chain for microphysical invigoration whereby greater aerosol concentrations increase atmospheric humidity, heating the troposphere as updrafts are cooled less by entrainment, and triggering large-scale ascent under WTG. Large-scale ascent imports moisture and enhances the initial humidity anomaly; differences in our simulations between humidity anomalies in RCE (Fig. 3) and WTG (Fig. S1) suggest that this feedback can amplify humidity perturbations by a factor of 4 or more. The strength of this feedback is difficult to estimate a priori and may be sensitive to details of the WTG implementation. However, we expect qualitatively similar results (ascent under increased aerosol loading) as long as the initial atmospheric heating produces large-scale ascent.

Our results emphasize the importance of interactions between clouds and their environment in determining the cloud-ensemble response to changes in cloud microphysics: the "humidity-entrainment" mechanism that invigorates convection in our simulations is intrinsically linked to the role that convective clouds play in determining environmental humidity.

A contemporaneous modeling study *(27)* also emphasizes cloud-environment feedbacks as an agent of microphysical invigoration. Like our study, theirs finds that higher $N_c$ increases cloud evaporation, which in turn invigorates convection without the need for ice-phase microphysics. Unlike our study, however, theirs focuses on changes in convection in RCE, which responds to very different constraints than convection in the WTG simulations that are the focus of our paper.

Our results stand in contrast to a recent study *(28)* that found only a weak link between aerosol concentrations and precipitation in a similar set of WTG-constrained simulations. The key difference is likely their focus on simulations with strong large-scale ascent forced by an SST perturbation. Analytical models and numerical sensitivity experiments (Supplementary Text and Fig. S3-S5) both suggest that humidity and precipitation become less sensitive to changes in detrainment (and thus to changes in aerosol concentrations) in the presence of strong large-scale ascent.

In the tropics, updraft speeds appear highest in regions with a moderately dry troposphere and not in regions where large-scale ascent pushes the atmosphere toward saturation *(7)*. Our numerical sensitivity experiments are consistent with this observation: average updraft speeds in strong convective cores are highest in simulations with weak ascent or moderate subsidence (Fig. S6). Because relative humidity is most sensitive to changes in detrainment when ascent is weak, the humidity-entrainment mechanism is likely to increase the frequency of strong updrafts most effectively in regions with a moderately dry troposphere and fast updrafts, although humidity increases may also lead to an accompanying decrease in peak updraft speeds.

Finally, because the humidity-entrainment mechanism relies on a three-way link between aerosols, humidity, and convective vigor—all of which are either directly or indirectly observable—our results provide a target for future observational work. Although the connections found in our idealized models may be challenging to detect in observations, doing so would point to a key pathway by which aerosols (including those produced by human activity) modify weather in the tropics.

**Acknowledgments:** We thank Xin Rong Chua, Yi Ming, and Raphael Rousseau-Rizzi for discussions about this work; Julia Wilcots and Henri Drake for comments on the manuscript; Marat Khairoutdinov for providing SAM; and Peter Blossey and an anonymous reviewer for constructive reviews. **Funding:** T.H.A. and T.W.C. were supported by NSF AGS 1740533: Convection and Rainfall Enhancement over Mountainous Tropical Islands. **Author contributions:** T.H.A. and T.W.C. designed research and wrote the paper; T.H.A performed research and analyzed data. **Competing interests:** The authors declare no competing interests. **Data and materials availability:** SAM is publicly available (http://rossby.msrc.sunysb.edu/~marat/SAM.html). Modifications to the SAM source code, simulation input files, and simulation output data underlying this work are archived at (*29*).

**List of Supplementary Materials:**

Materials and Methods

Supplementary Text

Figs. S1-S8

References (*30-38*)

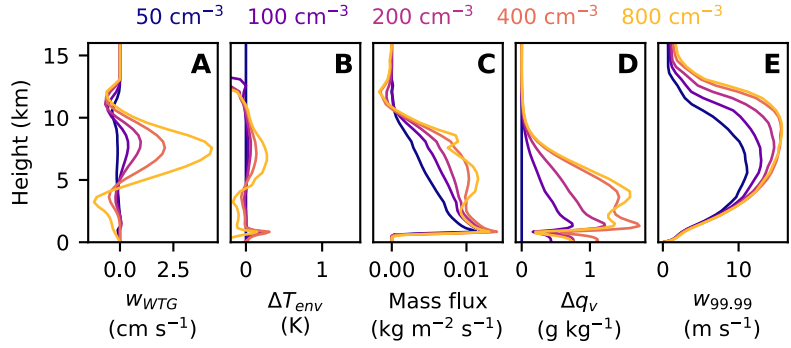

**Fig. 1.** Invigoration in WTG simulations. Panels (A-E) show WTG vertical velocities ($w_{WTG}$), domain- and time-mean temperatures ($T_{env}$), cloud mass fluxes, domain- and time-mean specific humidity ($q_v$), and 99.99$^{th}$ percentile vertical velocities ($w_{99.99}$) in WTG simulations with varied $N_c$. Domain- and time-mean temperatures and specific humidities are plotted as a difference ($\Delta$) from the WTG simulation with $N_c = 50$ cm$^{-3}$. Cloud mass fluxes include contributions from all grid points with cloud water mixing ratios above the smaller of $10^{-2}$ g kg$^{-1}$ or 1 percent of the saturation specific humidity.

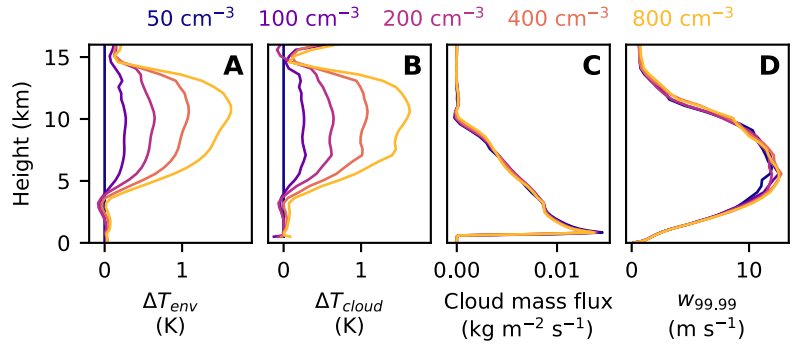

**Fig. 2.** Warming in RCE simulations. Panels (A-D) show domain- and time-mean temperatures ($T_{env}$), mean in-cloud temperatures ($T_{cloud}$), cloud mass fluxes, and 99.99[th] percentile vertical velocities ($w_{99.99}$) in RCE simulations with varied $N_c$. Temperatures are plotted as a difference ($\Delta$) from the WTG simulation with $N_c = 50$ cm[-3]. In-cloud temperatures and cloud mass fluxes include contributions from all grid points with cloud water mixing ratios above the smaller of 10[-2] g kg[-1] or 1% of the saturation specific humidity.

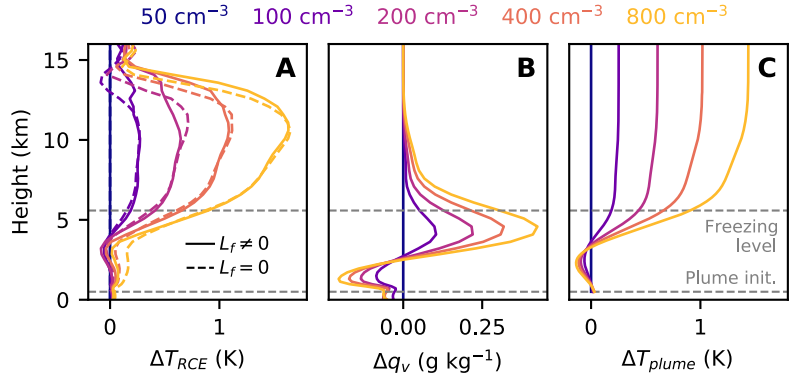

**Fig. 3.** RCE mechanism-denial experiments and plume calculations. Differences ($\Delta$) from an $N_c$ = 50 cm$^{-3}$ control of (A) domain- and time-mean temperature profiles from RCE simulations with default $L_f$ (solid lines) and $L_f$ = 0 (dashed lines), (B) domain- and time-mean specific humidity profiles from RCE simulations used as input for plume calculations, and (C) environmental temperature profiles from plume calculations. The gray dashed line at 500 m shows the level where plume calculations are initialized with domain- and time-mean temperatures from simulations, and the gray dashed line near 5 km indicates the freezing level, calculated as the lowest model level where more than 5% of the mean cloud water mass is ice.

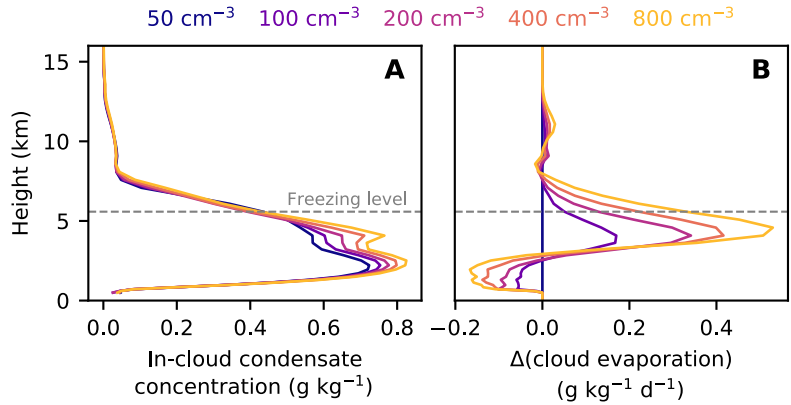

**Fig. 4.** In-cloud condensate concentration and cloud evaporation in RCE simulations. The in-cloud condensate concentration (A) is calculated as a conditional average over all grid cells where the liquid plus ice cloud mass concentration is larger than the smaller of $10^{-2}$ g kg$^{-1}$ and 1% of the saturation specific humidity. The cloud evaporation rate (B) is calculated as an average over all grid cells and plotted as a difference ($\Delta$) from the $N_c = 50$ cm$^{-3}$ simulation. The gray dashed line indicates the freezing level calculated as in Fig. 3.

# Supplementary Materials for

Aerosol invigoration of atmospheric convection through increases in humidity

Tristan H. Abbott, Timothy W. Cronin

Correspondence to: thabbott@mit.edu

**This PDF file includes:**

**Materials and Methods**

<u>Simulations</u>

Our simulations use version 6.11.3 of the System for Atmospheric Modeling (SAM) *(15)* integrated in a doubly-periodic non-rotating domain at a horizontal resolution of 1 km. The vertical grid contains 64 levels, with 12 levels in the lowest km, 500 m vertical spacing between 1 km and 20 km, and the model top at 27 km. Gravity waves are damped by a sponge layer in the top third of the domain.

The bottom boundary is a fixed-temperature ocean surface, set to 300 K except in WTG simulations with perturbed SST. Surface fluxes of latent and sensible heat are calculated using bulk aerodynamic formulae with exchange coefficients from Monin-Obukhov theory, and subgrid-scale fluxes in the atmosphere are parameterized using a first-order Smagorinsky scheme. Incoming solar radiation is set to an equinoctial diurnal cycle at 19.25 degrees from the equator, and radiative heating rates are calculated interactively using the Rapid Radiative Transfer Model *(30)* with greenhouse gas concentrations fixed at present-day levels. We initialize convection by adding a small amount of thermal noise near the surface, allow simulations to spin up for 40 days, and calculate statistics over the subsequent 20 days.

We run most of our simulations on a 128x128 km$^2$ domain. However, RCE mechanism denial experiments with the latent heat of fusion set to 0 self-aggregate in 128x128 km$^2$ domains. Because self-aggregation modifies environmental temperature profiles *(31)*, this complicates the comparison to RCE simulations with the default latent heat of fusion (which do not self-aggregate on 128x128 km$^2$ domains). Accordingly, the RCE simulations shown in Figs. 2-4 were run on a smaller 64x64 km$^2$ domain, which prevents the mechanism denial experiments from self-aggregating. The reduction in domain size has little effect on temperatures, humidities, mass fluxes, or updraft speeds in RCE simulations with the default latent heat of fusion (compare Fig. 2-3 with Fig. S7-S8). All WTG simulations, and the RCE simulation used to generate the WTG reference temperature profile, use 128x128 km$^2$ domains.

<u>Parameterizations of large-scale dynamics</u>

The WTG-based parameterization of large-scale dynamics is a "conventional WTG" scheme *(32)*. During each time step, the parameterization first diagnoses a large-scale vertical velocity based on differences between horizontally-averaged temperatures and a reference profile and then calculates large-scale tendencies by vertically advecting horizontal-average profiles. The vertical velocity $w_{WTG}$ is calculated as

$$w_{WTG} = \frac{1}{\tau_{WTG}} \sin\left(\frac{\pi z}{h}\right)\left(\frac{\bar{s} - s_0}{\partial_z \bar{s}}\right). \qquad (1)$$

Here, $s$ is dry static energy and $z$ is height. $s_0(z)$ is a reference dry static energy profile, which the parameterization takes as input, $\bar{s}(z)$ is the horizontal-average dry static energy profile, and $\partial_z \bar{s}$ is its vertical gradient (set to a minimum value of 0.3 J kg$^{-1}$ m$^{-1}$). The WTG relaxation timescale $\tau_{WTG}$ is calculated as the time required for a troposphere-deep gravity wave to propagate a distance $L_{WTG}$:

$$\tau_{WTG} = \frac{\pi L_{WTG}}{hN} . \qquad (2)$$

$N^2$ is the Brunt-Vaisala frequency (a measure of stability to vertical displacements) calculated from horizontal-average profiles of potential temperature $\overline{\theta}$ as

$$N^2 = \left\langle \frac{g}{\overline{\theta}} \frac{\partial \overline{\theta}}{\partial z} \right\rangle , \qquad (3)$$

where angle brackets indicate a vertical average between $z = 0$ and $z = h$ and $g$ is gravitational acceleration. The factor of $\sin(\pi z / h)$ is a mask that helps to produce vertical velocity profiles with structures similar to those found in the real tropical atmosphere, effectively making temperature relaxation timescales longer in the lower and upper troposphere, and the tropopause height $h$ is defined as the height at which domain-average temperature profiles first drop below 210 K. We set $L_{WTG}$ to 100 km, comparable to our simulation domain. Finally, we linearly taper $w_{WTG}$ to 0 at the surface starting from a height of 1 km.

We then use $w_{WTG}$ to calculate source and sink terms for model fields by vertically advecting horizontal-average model fields. For a generic model field $\phi$ this gives a source term

$$\dot{\phi}_{WTG}(z) = -w_{WTG} \frac{\partial \overline{\phi}}{\partial z} . \qquad (4)$$

We add source terms for liquid/ice water static energy (the prognostic thermodynamic variable used by SAM) and for all variables that represent mass concentrations of different water species (including water vapor, both phases of cloud water, and all classes of precipitating water).

The net effect of the WTG-based parameterization is to remove horizontal temperature differences relative to a reference state defined by $s_0$ and, additionally, provide sources and sinks of moisture that are consistent with the removal of horizontal temperature gradients. In all WTG simulations, we use $s_0(z) = c_p T_0(z) + gz$, where $c_p$ is the heat capacity of dry air and $T_0(z)$ is the time- and domain-average temperature profile from the last 20 days of an RCE simulation on a 128x128 km$^2$ domain with $N_c = 50$ cm$^{-3}$.

Entraining plume calculations for RCE simulations

Following *(24)*, the entraining plume calculation used in Fig. 3 assumes that the plume buoyancy is small, allowing us to write an equation for the evolution of plume moist static energy (MSE, $h$) with height $z$ as

$$\frac{dh}{dz} = -\varepsilon L_v (q^* - q_v) . \qquad (5)$$

Here, $\varepsilon$ and $q^*$ are the plume entrainment rate and the in-plume saturation specific humidity, $q_v$ is the specific humidity of the environment surrounding this plume, and $L_v$ is the latent heat of vaporization. To solve this model, we integrate $h$ upward from 500 m using the environmental temperature at 500 m from simulations as the initial condition. We diagnose $q^*$ from $h$ during the integration by assuming saturation in the plume and set $q_v$ equal to domain- and time-mean water vapor profiles from simulations.

**Supplementary Text**

<u>Relationship between cloud mass flux and WTG vertical velocity</u>

Clear-air energy balance and mass continuity can be combined to argue that, in WTG simulations, increases in the WTG vertical velocity should be accompanied by increases in the simulated cloud mass flux. During this discussion, we use "resolved" to refer to flows simulated explicitly by the CRM rather than parameterized by the WTG scheme.

In steady state, and ignoring the effects of precipitation evaporation, clear-air energy balance requires that radiative cooling ($Q$) be balanced by a combination of warming by resolved subsidence ($w_{sub}$, positive downward) and cooling by the WTG vertical velocity ($w_{WTG}$, positive upward):

$$\left(w_{sub} - w_{WTG}\right)\frac{\partial s}{\partial z} = Q , \qquad (6)$$

where $s$ is the dry static energy (whose profile is fixed under WTG). Mass continuity within the sub-domain-scale flow simulated by the CRM requires that the resolved upward cloud mass flux ($M_{res}$) be balanced by resolved subsidence in clear air, or that

$$M_{res} = \rho w_{sub}(1 - f) , \qquad (7)$$

where $\rho$ is density and $f$ is the fractional area covered by cloud. The WTG vertical velocity contributes an additional implied upward cloud mass flux $M_{WTG}$ of

$$M_{WTG} = \rho w_{WTG} f , \qquad (8)$$

but $M_{WTG}$ is likely small compared to $M_{res}$ under normal circumstances, since typically the cloud fraction $f << 1$, and $Q > 0$ (i.e., radiation cools the atmosphere) requires that $w_{sub} > w_{WTG}$. Defining the net clear-air subsidence speed required to balance radiative cooling as $w_e = w_{sub} - w_{WTG}$ and assuming that the cloud fraction is small gives

$$M_{res} \approx \rho\left(w_{WTG} + w_e\right) . \qquad (9)$$

This result indicates that, absent significant changes in radiative cooling, increases in WTG vertical velocity are likely to be accompanied by increases in the simulated cloud mass flux. Considering condensate evaporation is unlikely to qualitatively change this result: if cooling from condensate evaporation increases with large-scale ascent (larger $w_{WTG}$) due to increases in precipitation and is balanced by faster clear-air subsidence (larger $w_e$), then both terms in Eq. 9 promote an increase in cloud mass flux with stronger large-scale ascent.

<u>Scalings for changes in parcel temperatures</u>

Using $\Delta$ to denote differences between simulations with different microphysical properties, changing the amount of frozen cloud condensate by $\Delta q_{c,freeze}$ leads to a change in parcel temperature $\Delta T_{freeze}$ of

$$\Delta T_{freeze} = \frac{L_f}{c_p} \Delta q_{c,freeze} , \qquad (10)$$

where $c_p$ is the heat capacity of air and $L_f$ is the latent heat of fusion. Deep convective clouds have peak cloud water concentrations of around 3 g kg$^{-1}$ *(33-34)*. This upper bound on $\Delta q_{c,freeze}$ gives

$$\Delta T_{freeze} \lesssim 1 \text{ K}; \qquad (11)$$

actual values of $\Delta q_{c,freeze}$ and $\Delta T_{freeze}$ are likely to be substantially smaller.

A scaling for the impact of changes in humidity can be derived starting from Eq. 4 of *(24)*, which uses a zero-buoyancy plume model to provide an expression for the difference between the temperatures of an entraining plume ($T_{entrain}$) and an undilute moist adiabat ($T_{undilute}$):

$$T_{entrain} - T_{undilute} = -\frac{L_v}{c_p + \dfrac{L_v^2 q^*}{R_v T^2}} \varepsilon (1 - RH) \int_0^z q^*(z')dz' , \qquad (12)$$

where $R_v$ is the specific gas constant for water vapor, $L_v$ is the latent heat of vaporization, $\varepsilon$ is an entrainment rate, $q^*$ is saturation specific humidity, and $RH$ is relative humidity. If we now consider two entraining plumes that have identical properties at cloud base ($z = 0$) but rise through environments with relative humidities that differ by $\Delta RH$, we can write the temperature difference between the two plumes $\Delta T_{entrain}$ as

$$\Delta T_{entrain} = \frac{L_v}{c_p + \dfrac{L_v^2 q^*}{R_v T^2}} \varepsilon \Delta RH \int_0^z q^*(z')dz' . \qquad (13)$$

The temperature difference is largest high in the atmosphere, where

$$\Delta T_{entrain} \approx \frac{L_v}{c_p} \varepsilon \Delta RH \int_0^\infty q^*(z')dz' . \qquad (14)$$

Approximating $q^*(z)$ as an exponential with a decay height of $H_q = 2$ km allows us to re-write $\Delta T_{entrain}$ as

$$\Delta T_{entrain} \approx \frac{L_v}{c_p} \varepsilon H_q q_s^* \Delta RH . \qquad (15)$$

If we use $q_s^* \approx 20$ g kg$^{-1}$ (corresponding to a surface temperature of about 300 K) and an entrainment rate of 1 km$^{-1}$ *(35)*, this gives

$$\Delta T_{entrain} \approx 100 \times \Delta RH . \qquad (16)$$

This suggests that an increase in relative humidity of just 1% can change parcel temperatures by ~1 K.

The coefficient multiplying relative humidity changes in Eq. 16 is sensitive to assumptions about entrainment. Although the entrainment rate we use is consistent with measurements in CRMs (*35*) and plume calculations that provide a good fit to our model results (Fig. 3), choosing a lower entrainment rate would reduce the apparent sensitivity to relative humidity changes. For example, Fig. 4 of (*36*) displays good agreement between temperature profiles from CRM simulations and an entraining plume model with an entrainment rate of $0.8/z$ (where $z$ is height) and shows that this entrainment profile produces temperature increases of about 0.3 to 0.4 K per 1% change in relative humidity. (Buoyancy changes, estimated by digitizing the plot, peak at 0.15 to 0.2 m s$^{-2}$ per 10% change in relative humidity at heights of about 12 km, corresponding to temperature changes of about 3-4 K assuming an absolute temperature of 200 K.) Despite uncertainties arising from poorly-constrained entrainment rates, however, these estimates broadly suggest that parcel temperatures are likely to be much more sensitive to changes in environmental humidity than to changes in cloud water concentrations.

Sensitivity of aerosol invigoration to large-scale forcing

Simple analytic models give expressions for tropical relative humidity of the form

$$RH = \frac{\delta_T}{\delta_T + \frac{M_d}{M_u}\gamma},$$ (17)

where $\delta_T$ is a fractional detrainment rate per unit distance (dimensions of inverse length) that represents moistening by mixing saturated air and condensate from clouds into the environment, $\gamma = -\partial_z \ln q^*$ represents the inverse of the length scale over which subsidence dries the atmosphere, $M_d$ is the downward clear-air mass flux, and $M_u$ is the in-cloud updraft mass flux (*26*).

For reasonable values of $\delta_T$, $M_d/M_u$, and $\gamma$, Eq. 17 implies that relative humidity becomes less sensitive to changes in detrainment moistening under the influence of large-scale upward motion. More specifically, differentiating Eq. 17 and multiplying by $\delta_T$ gives an expression for the sensitivity of relative humidity to fractional changes in the detrainment rate:

$$\frac{\partial RH}{\partial \ln \delta_T} = \frac{\delta_T \frac{M_d}{M_u}\gamma}{\left(\delta_T + \frac{M_d}{M_u}\gamma\right)^2}.$$ (18)

In turn, this can be re-written in terms of the RCE relative humidity

$$RH^* = \frac{\delta_T}{\delta_T + \gamma}$$ (19)

as

$$\frac{\partial RH}{\partial \ln \delta_T} = \frac{\frac{M_d}{M_u}\frac{RH^*}{1-RH^*}}{\left(\frac{M_d}{M_u} + \frac{RH^*}{1-RH^*}\right)^2}.$$ (20)

The sensitivity is largest when $M_d/M_u = RH^*/(1-RH^*)$. Near RCE, where large-scale vertical motion is weak and $M_d/M_u \approx 1$, tropospheric relative humidity is typically 0.6 to 0.8 (*25*). This corresponds to $RH^* / (1 - RH^*) \approx 1.5$ to $4$, which is larger than $M_d / M_u$. As a result, the sensitivity of relative humidity to detrainment should be smaller in the presence of large-scale ascent ($M_d / M_u < 1$) than in RCE ($M_d / M_u = 1$). The sensitivity only increases moving away from RCE into a regime with mean ascent if $RH^* < 0.5$, much smaller than is typical in RCE (Fig. S3).

If changes in aerosol concentration influence tropospheric humidity primarily by altering the detrainment of condensed water ($\delta_r$), Eq. 20 also implies that relative humidity should become less sensitive to $N_c$ as large-scale ascent increases. We test this prediction in two ways: first, by performing additional RCE simulations with an imposed large-scale vertical velocity profile $w_{LS}$ given by

$$w_{LS} = \begin{cases} w_0 \sin(\pi z / H) & z \leq H \\ 0 & z > H \end{cases} \qquad (21)$$

with $H = 13$ km, similar to *(26)*; and second, by using a set of WTG simulations where large-scale motion is induced by changing SST. For reference, Fig. S3 includes mass flux ratios at 5 km for RCE simulations with large-scale ascent, diagnosed following *(26)* as

$$\frac{M_d}{M_u} = \frac{M_{res} - \rho w_{LS}}{M_{res}}, \qquad (22)$$

where $M_{res}$ is the resolved cloud mass flux, $\rho$ is density, and $w_{LS}$ is the large-scale vertical velocity (Eq. 21). (Note that because mass is conserved separately for resolved and large-scale flows in our simulations, using Eq. 22 results in diagnosing negative mass flux ratios in some simulations with strong large-scale ascent. This likely corresponds to a limit where one of the assumptions in the plume model of *(26)*---namely, that humidity and subsidence speeds are uniform and constant outside of clouds---breaks down.)

Because Eq. 20 provides an estimate of the sensitivity of humidity to changes in detrainment holding vertical motion fixed, it only provides a direct estimate of humidity changes in RCE simulations with imposed vertical motion. However, because humidity changes under WTG occur through an initial perturbation from aerosols plus a feedback from changes in large-scale ascent, we also expect the total WTG humidity change to be small when the sensitivity of humidity to changes in detrainment alone is weak. We include WTG simulations to test whether this expectation is borne out. Large-scale source terms are calculated identically in WTG simulations and in RCE simulations with prescribed large-scale vertical motion, except that the RCE simulations calculate source terms by vertically advecting horizontally-resolved fields and include source terms for horizontal momenta and number concentrations of condensed water species that are neglected in the WTG simulations. Like all WTG simulations, the RCE simulations with prescribed vertical motion use a 128x128 km$^2$ domain.

Our WTG simulations appear to be only marginally stable to the development of large-scale circulations (consistent with bistability found in previous modeling studies *(37-38)*) and small changes in SST can easily push them into moist regimes with strong ascent or dry regimes with strong subsidence and no deep convection (Fig. S4). The transition to a moist regime as SST increases seems to result from a jump between distinct equilibria rather than a gradual shift of a single equilibrium: a small increase in SST pushes the simulations into a limit cycle where domain-average precipitation rates vary between about 5 and 30 mm day$^{-1}$, and larger SST increases allow the simulations to remain in the high-precipitation state. Similarly, simulations with decreased SST either remain in a state with precipitation rates near that of the RCE simulations or jump to a dry state with no precipitation.

Although these regime transitions make it difficult to produce gradual changes in large-scale vertical motion by gradually changing SST, we can nevertheless contrast the impact of aerosol perturbations in three regimes: one RCE-like with weak vertical motion (including most simulations with SST perturbations less than about 0.25 K in magnitude), one with strong upward motion (simulations with SST perturbations of ~0.5 K and above), and one with strong subsidence (SST perturbations of ~-0.5 K and below).

As $w_0$ increases from 0 to 1 and then 5 cm s$^{-1}$ in RCE simulations with imposed vertical motion, increases in $N_c$ have a progressively smaller impact on relative humidity (Fig. S5A), consistent with higher $N_c$ promoting higher relative humidity by increasing detrainment moistening below the freezing level. Similarly, inducing large-scale ascent by increasing SST in WTG simulations weakens the sensitivity of relative humidity to $N_c$ (Fig. S5B), and sensitivity to $N_c$ is largest in regimes with weak subsidence (small SST decreases). Precipitation becomes insensitive to changes in Nc for large positive SST perturbations (Fig. S5C), consistent with the results of (**28**).

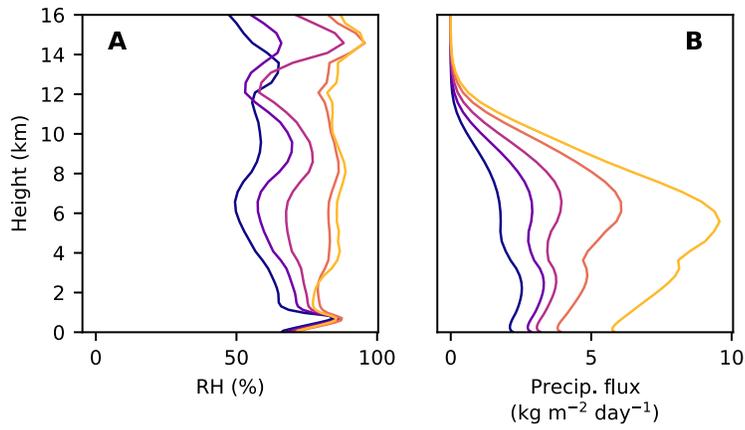

**Fig. S1.** Relative humidity (RH, A) and downward precipitation flux (B) in WTG simulations. Relative humidity is calculated using a linear combination of saturation specific humidity over liquid water and ice, weighted by the fraction of cloud condensate in each phase at each level.

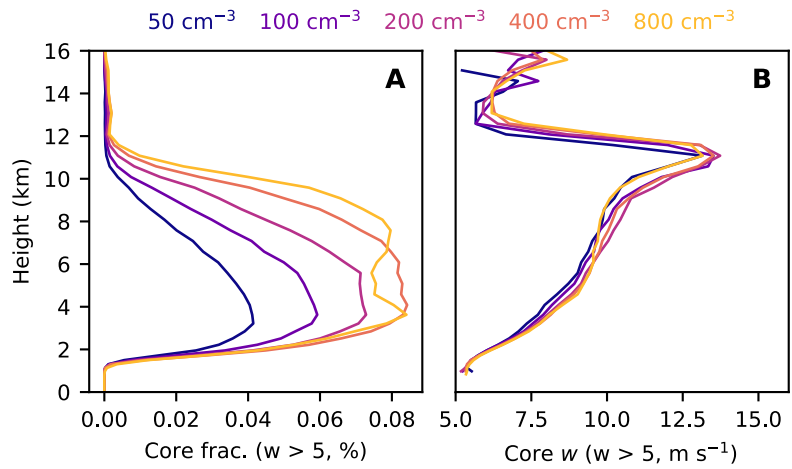

**Fig. S2.** Convective cores in WTG simulations. Panel (A) shows the areal fraction occupied by strong convective cores (defined as gridpoints with positive buoyancy and vertical velocity above 5 m s⁻¹), and panel (B) shows the average vertical velocity within strong convective cores.

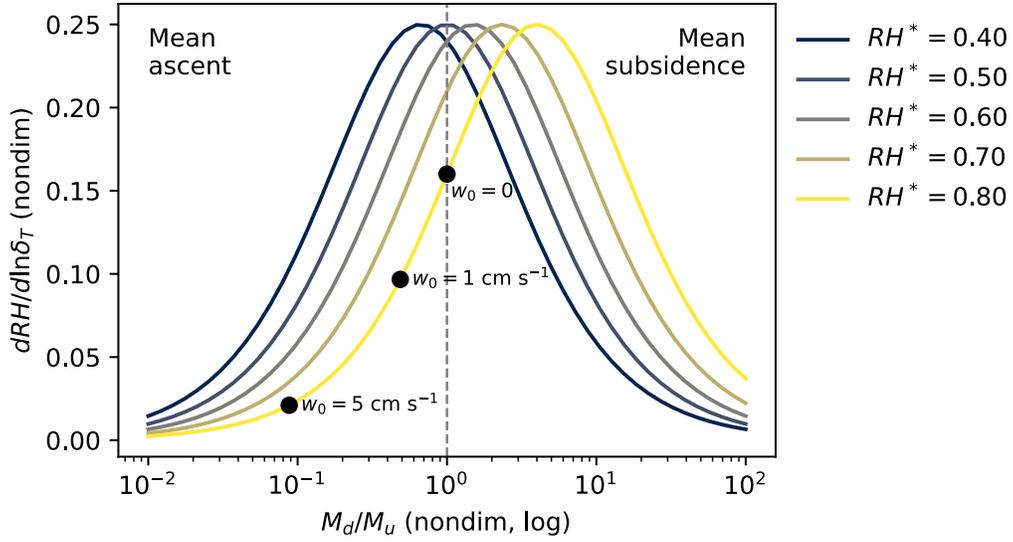

**Fig. S3.** Relative humidity-detrainment relationship in a simple model. Sensitivity of relative humidity to changes in the effective detrainment rate (Eq. 15) for different values of RCE relative humidity $RH^*$. The gray dashed line indicates $M_d/M_u = 1$ (i.e., RCE), and upward motion increases moving to the left. Black dots indicate mass flux ratios diagnosed at 5 km from RCE simulations with imposed large-scale ascent (details in text), plotted for convenience with vertical locations corresponding to the sensitivity curve for $RH^* = 0.8$.

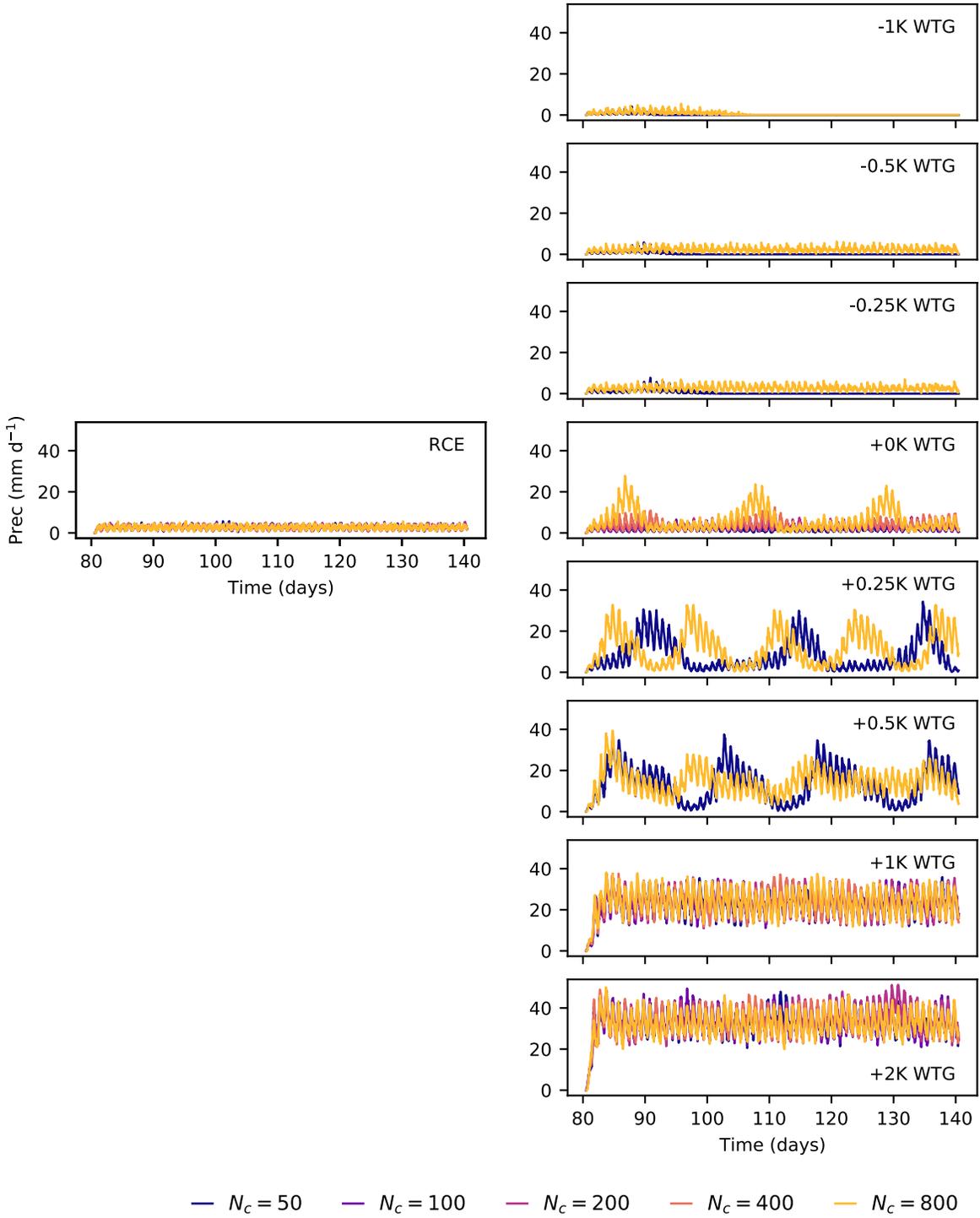

**Fig. S4.** Precipitation timeseries in WTG SST perturbation experiments. Domain-average precipitation timeseries in (top row) WTG simulations with SST increases and (bottom row) RCE simulations and WTG simulations with SST decreases. All panels include simulations with $N_c = 50$ cm$^{-3}$ and $N_c = 800$ cm$^{-3}$, but not all include intermediate $N_c$.

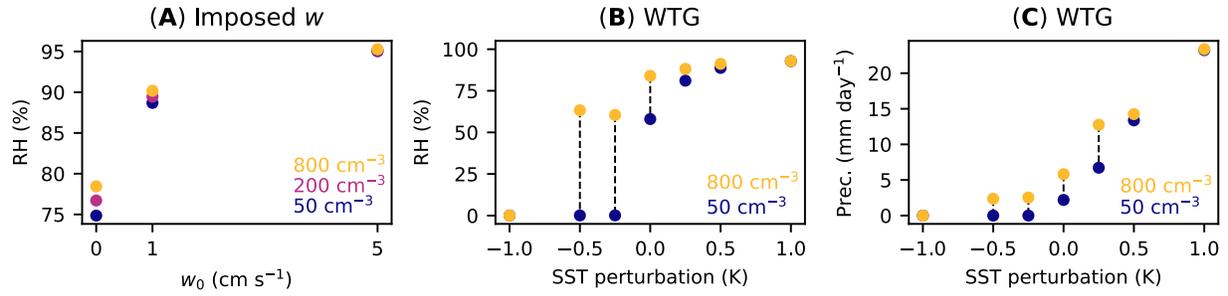

**Fig. S5.** Humidity and precipitation in simulations with large-scale vertical motion. (A) 2-6 km relative humidity in RCE simulations with imposed vertical motion, calculated using a linear combination of saturation vapor pressures over liquid and ice weighted by the fraction of cloud water mass in each phase and a mass-weighted vertical average. (B,C) 2-6 km relative humidity, calculated as in panel (A), and surface precipitation from WTG-constrained simulations. Black dashed lines connect simulations that are identical except for the liquid cloud droplet number concentration.

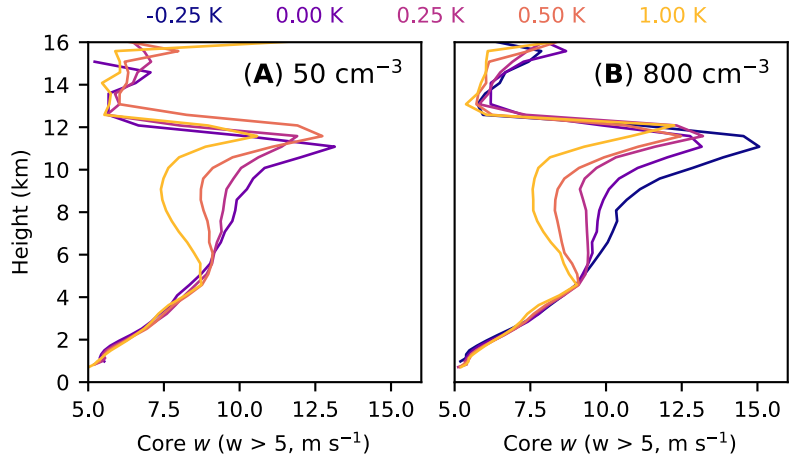

**Fig. S6.** Convective core updraft speeds in WTG SST perturbation experiments. Both panels show average updraft speeds within strong convective cores as a function of SST perturbation; profiles in panel (A) are from simulations with $N_c = 50$ cm$^{-3}$, and profiles in panel (B) are from simulations with $N_c = 800$ cm$^{-3}$. Strong convective cores are defined as in Fig. S2 as grid points with positive buoyancy and vertical velocity above 5 m s$^{-1}$. The $N_c = 50$ cm$^{-3}$ simulation with a -0.25 K SST perturbation contains no grid points that qualify as strong convective cores.

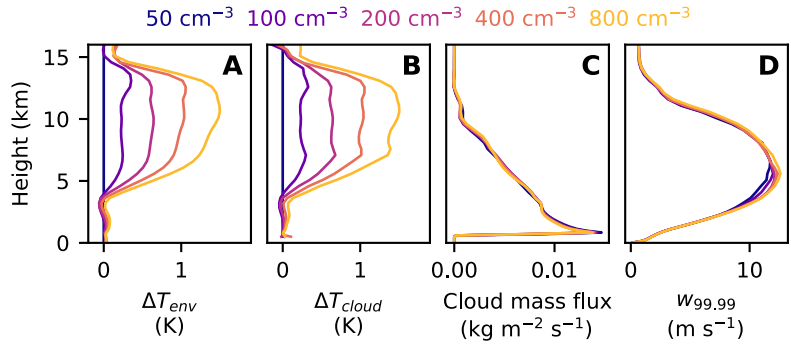

**Fig. S7.** As in Fig. 2 in the main text, but for RCE simulations on a 128x128 km$^2$ domain. Panels (A-D) domain- and time-mean temperatures ($T_{env}$), mean in-cloud temperatures ($T_{cloud}$), cloud mass fluxes, and 99.99[th] percentile vertical velocities ($w_{99.99}$) in RCE simulations with varied $N_c$. Temperatures are plotted as a difference ($\Delta$) from the WTG simulation with $N_c$ = 50 cm$^{-3}$.

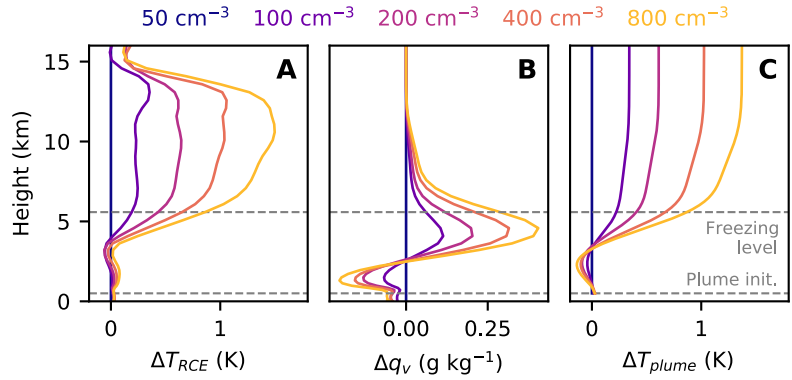

**Fig. S8.** As in Fig. 3 in the main text, but for RCE simulations in a 128x128 km² domain. Panels (A-B) show differences from a control RCE simulation (with $N_c = 50$ cm⁻³) in (A) domain- and time-mean temperature profiles and (B) domain- and time-mean specific humidity profiles from RCE simulations used as input for plume calculations. Panel (C) shows environmental temperature profiles from plume calculations. The gray dashed line at 500 m shows the level where plume calculations are initialized with domain- and time-mean temperatures from simulations, and the gray dashed line near 5 km indicates the freezing level, calculated as the lowest model level where more than 5% of the mean cloud water mass is ice.